Isotopic Measurements of SNM using a Portable Neutron Resonance Transmission System for Arms Control

Mital A. Zalavadia[1], Ethan A. Klein[2], Michael E. Moore[1], Jonathan A. Kulisek[1], Farheen Naqvi[2], Glen A. Warren[1], Areg Danagoulian[2]

[1]Pacific Northwest National Laboratory, Richland, Washington 99354, USA
[2]Department of Nuclear Science and Engineering, MIT, Cambridge, Massachusetts 02139, USA

**Abstract:** Neutron Resonance Transmission Analysis (NRTA) was explored as an arms control verification approach to support potential future nuclear weapon limiting treaties. A compact and portable neutron Time of Flight (ToF) system was developed to enable proof-of-concept NRTA measurements of special nuclear material (SNM). Using a short 2-meter flight path, the NRT system is sensitive to cross-section resonances of isotopes such as $^{235}$U, $^{238}$U, $^{239}$Pu and $^{240}$Pu between 1-100 eV incident neutron energies due to their physical nuclear structure. The detected neutron ToF spectrum exhibits transmission dips at resonance energies that are characteristic of SNM isotopic composition in the inspected item. The proof-of-concept measurements of Highly Enriched Uranium (HEU), Depleted Uranium (DU), and Reactor Grade Plutonium (RGPu) confirmed the characteristic resonance features within two hours of data collection time. Analysis via the REFIT resonance fitting tool accurately predicted the $^{235}$U enrichment and Pu isotopic composition within 5% and 6% of the known values, respectively.

## 1.0 INTRODUCTION AND BACKGROUND

Potential future international agreements related to nuclear warhead limitations may necessitate verification of warhead components throughout their lifecycle, e.g., confirming various attributes of Special Nuclear Material (SNM) prior to and following dismantlement. A range of concepts and approaches exists to address such a need [1-3]. State-of-the-art systems aim to build confidence by confirming templates or attributes of Treaty Accountable Items (TAIs) using passive gamma spectra; the Trusted Radiation Identification System (TRIS) [1, 3] and the Trusted Radiation Attribute Demonstration System (TRADS) [2] are the respective typically cited examples. As gamma-ray-based systems, the presence of high-Z materials and self-attenuation impacts the observable confirmation signatures. Confirmation of Highly Enriched Uranium (HEU) continues to be a challenge for similar passive gamma-ray confirmation-based techniques due to the low energy of gamma emissions and the strong self-attenuation of uranium. Passive neutron techniques also prove challenging due to weak spontaneous fission neutron intensities for $^{235}$U and $^{238}$U [4]. Therefore, active Non-Destructive Assay (NDA) techniques may provide more confidence for verification measurements of TAIs containing HEU.

This effort investigated the use of a portable Neutron Resonance Transmission Analysis (NRTA) system as a potential means to verify a TAI by confirming its isotopic composition. NRTA is based on measuring transmitted neutron energy by recording the neutron time-of-flight (ToF), i.e., the time the neutron took to travel from the source to the detector. High temporal resolution NRTA measurements are performed at beam line user facilities due to their precision beam timing and excellent beam linearity, typically expressed as a beam length to beam aperture diameter ratio, *L/d*. Large *L/d* values minimize the ToF uncertainty in the detected neutron energies. However, such a large setup is not practical for arms control verification exercises. Recent research has shown the potential of using a commercial deuterium-tritium (DT) neutron generator and short flight paths, < 5 m, to perform NRTA [5-7]. This effort builds upon that work and explored a compact 2-meter flight path system, with sufficient temporal resolution to be sensitive to 1-50 eV incident neutron resonances in SNM (see Figure 1). Previous studies explored the use of non-SNM targets, such as tungsten plates, to investigate the NRTA technique on shorter flight paths. Building on this foundation, the current work enhanced the NRTA methods by utilizing a neutron generator with a shorter pulse width. The technique was then applied to Pu, highly enriched uranium (HEU), and depleted uranium (DU) targets to measure and study the signatures relevant to arms control application. Additionally, the epithermal (~1 – 100 eV) neutron yield was increased by nearly 50% through improvements to moderator design. Modular neutron shield and collimator were developed to further reduce neutron background and increase system portability.

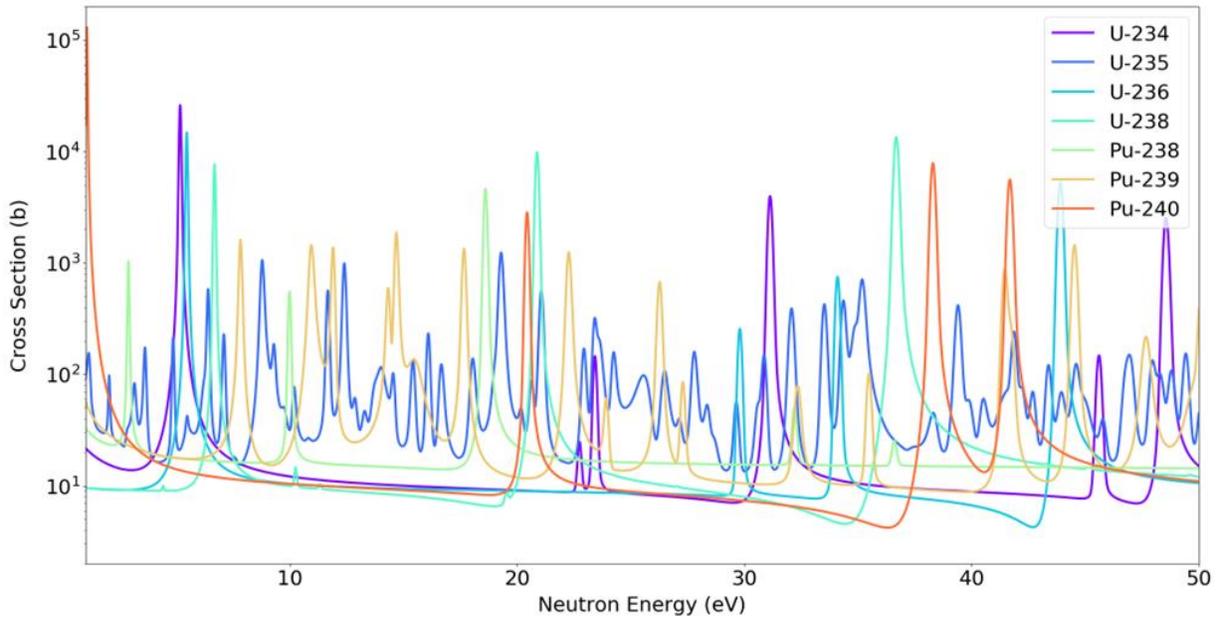

Figure 1: Resonance structure in the total neutron cross sections for U and Pu isotopes from 1 to 50 eV. Note the logarithmic scale for the cross section. Data from [8].

## 2.0	EXPERIMENTAL SETUP

The experimental NRTA system consisted of a moderated deuterium-tritium (D-T) neutron source, multiplier, moderator, filter, collimator, detector, and shielding (See Figure 2). The design aspects of each of the major components are discussed below.

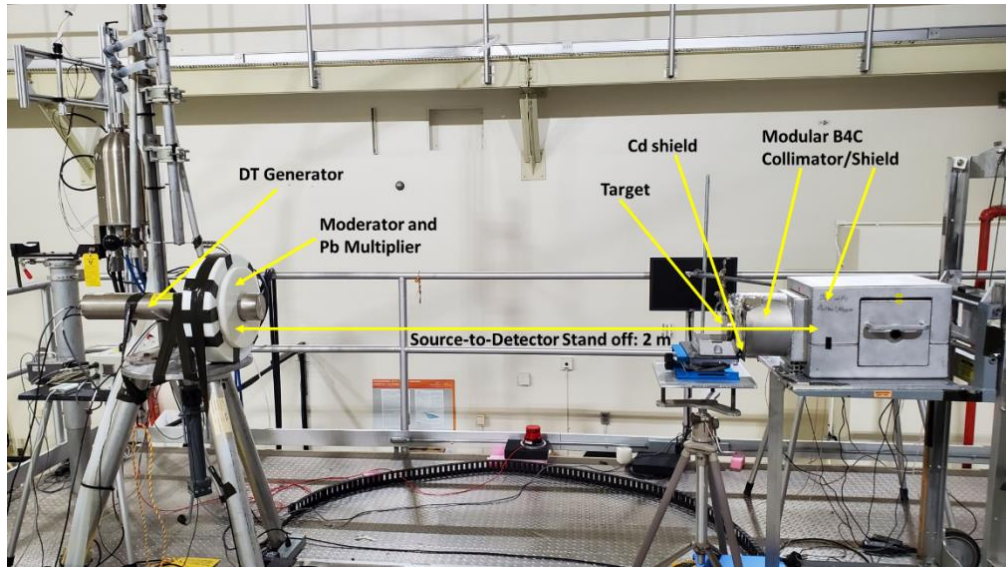

Figure 2: The Time-of-Flight Measurement System at PNNL's Low Scatter Facility.

## 2.1 Neutron Source

The neutron source needed to be portable, provide adequate neutron output, and be able to provide relatively fast pulses with a width of less than 1 µs for the envisioned measurement scenario. The neutron source system consisted of a Deuterium-Tritium (DT) neutron generator, a Pb multiplier, and HDPE moderator. There exists a number of commercial options for portable neutron generators providing a range of neutron output rate and timing characteristics [9]. The pulsed Thermo Fisher P-385 neutron generator was selected due to its ability to generate pulses down to 5 µs. The neutron yield of this generator was approximately $5 \times 10^8$ neutrons per second.

The time structure of the neutron pulse affects the overall system energy resolution. A wider neutron pulse prolongs the output of the neutrons over time, thereby resulting in a higher uncertainty in the $t_0$ and the ToF. The generator has a maximum acceleration voltage of 130 kV, a maximum beam current of 70 µA, and a minimum nominal duty cycle of 5.0%. The neutron pulse time structure was characterized at various beam currents, accelerator voltages, and duty cycle to determine the shortest possible pulse while maintaining stable generator operation. While the NRTA used GS20 for the detection of epithermal neutrons, for the neutron source characterization purposes the 14.1 MeV neutrons were detected using an EJ-309 liquid organic scintillator detector. The standoff from the source target plane on the generator tube to the detector was set to 2 meters. The analog signals from the PMT were read out using a CAEN 5730SB 14-bit 500 MS/S digitizer. The measurement durations were 100 seconds long. The CAEN CoMPASS utility was used to set appropriate gate lengths and to perform pulse shape discrimination (PSD) in real time. The time-of-flight spectrum was calculated in post-processing using the time difference between the time of the reference trigger signal from the generator and the time of detection for each detected neutron event.

The generator pulse frequency was set to 5000 Hz (200 µs period) for the optimization measurements. This frequency was primarily selected to allow adequate time for neutrons generated and moderated in the $i^{th}$ generation to make it to the detector 200 cm away before the next pulse, $i^{th} + 1$, is fired by the generator. This process of neutrons detected from i$^{th}$ + 1 generation prior to neutrons from i$^{th}$ generation are detected is referred to as wrap-around neutrons. Wrap-around neutrons can add to unwanted background and erroneous time-of-flight calculations. The frequency of 5000 Hz, and the associated period, ensures that a vast majority of neutrons in ~1-100 eV range are detected before wrap-around occurs. The contribution of the wrap-around neutrons was removed by the placement of a 2mm thick cadmium filter, which is a strong absorber for thermal neutrons with energies less than approximately 0.5 eV. A series of measurements was conducted at varying accelerator voltages and beam currents to arrive at a peak shape that is as close as possible to a Gaussian distribution. The duty factor was then optimized at these settings to obtain the shortest possible pulse at which the generator would operate in a stable regime.

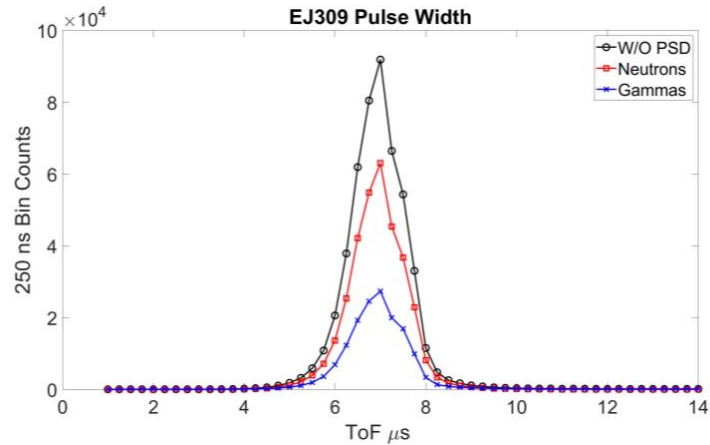

Figure 3: Optimized P-385 neutron pulse structure. The standard deviation of the pulse is approximately 0.6 μs.

Figure 3 shows the shortest pulse achieved by reducing the beam current and duty factor to 35 μA and 3.5%, respectively. At these settings, the pulse is much tighter and nearly Gaussian, with a sigma of 0.6 μs. The final generator settings for subsequent experiments were 130 kV accelerator voltage, 35 μA beam current, and 3.5 % duty factor.

*Neutron Moderator and Multiplier*

The 14.1 MeV monoenergetic neutrons from the generator need to be moderated to the epithermal range (i.e., ~1 – 100 eV) for the NRTA application envisioned and described earlier. Several aspects considered for the moderator design included axial epithermal neutron production, gamma-ray production, off-axis neutron production, moderation time, moderation distance, geometry, and material. These aspects are briefly discussed here. For neutrons to constitute a valid NRTA signal, it must have the required epithermal energy as it travels along the ToF axis and be detected in a reasonable amount of time given its energy. Monte Carlo N-Particle (MCNP) [10] simulations were conducted to track neutrons that were incident upon the front face of the moderator assembly with the energy in the 1-100 eV range. The neutrons were considered axial if the radial distance with respect to the ToF axis was less than 10 cm at 2 meters, where the detector is located. The vast majority of the axial, epithermal neutrons leave the moderator very close to the generator tube, therefore the reduction in the number of epithermal neutrons due to the presence of the detector collimator on the Monte Carlo tally surface is not expected to be significant. The neutrons interacting in the moderating material results in the production of gamma rays via thermal neutron capture. For hydrogenous materials, the $^1$H(n, $\gamma$)$^2$H reaction produces a 2.2 MeV gamma. The potential to utilize neutron-absorbing materials, such as boron or enriched lithium was considered to reduce potential gamma background. Off-axis neutrons arise when neutrons thermalize in the surrounding materials or in the vicinity of the detector. Given significant neutron backgrounds, an approach to shielding the detector was explored. The uncertainty in time it takes for a neutron to moderate can be accounted for in terms of an equivalent distance, which is dependent on the mean free path in the moderator. Therefore, the distribution in moderation time is not expected to change significantly with small variations in geometry. The spatial uncertainty exists due to the nonzero distance travelled by neutrons in the radial direction following moderation. Therefore, the effective ToF must count for the off-axis distance by summing this distance with the ideal ToF distance in quadrature: $d_{TOF}^{eff} = \sqrt{d_{rad}^2 + d_{TOF}^2}$. The thickness and the outer radius of the moderator limit the off-axis distance and, for relatively short time of flight distance, the impact of the spread on system energy resolution is expected to be negligible.

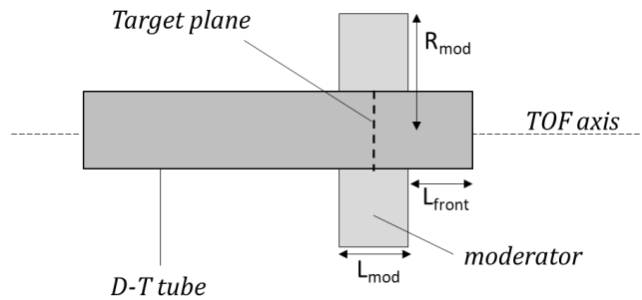

Figure 4: Moderator schematic and dimensional parameters

The neutron source design and optimization were largely based on the prior NRTA work by Klein et. al. [6]. Based on prior MCNP modeling, the general geometry considered for further optimizing consisted of an annulus placed around the target plane of the D-T tube aligned in an axial configuration with respect to the flight axis. The optimized parameters included material, axial and radial dimensions of the moderator, and axial position of the moderator with respect to the generator target plane. Figure 4 shows the moderator schematic and associated dimensional parameters.

Initially, both HDPE and heavy water (D$_2$O) were considered for neutron moderation. Both materials have low-Z, resulting in effective moderation of the 14.1-MeV neutrons to epithermal energies. The key differences between the two in their homogeneous forms are the moderating power and moderating ratio. A higher moderating power requires fewer collisions to moderate from fast to epithermal energies and thereby decreases the equivalent moderation delay distance subsequently reducing uncertainty in total time-of-flight distance and measured neutron energy. The moderating ratio takes into account the competing processes of neutron scattering vs. neutron absorption during the moderation process. Materials with higher relative neutron absorption cross-section not only eliminate the usable epithermal neutrons but also produce 2.2 MeV gammas upon neutron capture (i.e., prompt gammas), which can contribute to unwanted background.

Loading the HDPE moderator with materials having higher neutron capture cross-sections (i.e., neutron poisons) than hydrogen was considered as a way of reducing the gamma background arising from thermal neutron capture. While capture on hydrogen produces 2.2 MeV gammas, the neutron poisons either produce no gammas (in the case of $^6$Li) during neutron capture, or in the case of boron produce 460 keV gammas – which are easier to both shield and to filter out during the processing of data from GS20 scintillator. Commercially available boron and $^6$Li doped HDPE materials were modeled in MCNP to understand the potential benefits of loaded HDPE. Borated PE (polyethylene) results in a significant reduction in gammas while also significantly reducing the neutron flux. Based on the results, lithiated polyethylene commercially available at 7.5% weight fraction was found to be ideal in terms of reducing gamma background and maintaining epithermal flux. However, this commercial option has a hydrogen density of only two-thirds that of HDPE. Nevertheless, the MCNP simulations showed that the 2.2 MeV gammas do not contribute to significant background. Therefore, it was concluded that pure polyethylene (HDPE) should be the choice of moderator for further optimizing geometry.

MCNP simulations were performed to determine the optimal width ($R_{mod}$) thickness ($L_{mod}$), and position along the generator axis with respect to the target plane by varying these parameters to determine their impact on epithermal flux. The moderator of constant width $R_{mod}$ of 20 cm was centered around the target plane and the moderator thickness was varied by changing the back and front offset of the moderator with respect to the target plane. Changing the back and front offset effectively changes the overall thickness $L_{mod}$ of the moderator. An optimization study showed that the optimal moderator thickness is 5 cm in the front and 6 cm at the back relative to the target plane, giving the overall thickness of $L_{mod}$ = 11 cm. Given that HDPE slabs are commonly available in 1-inch thicknesses, ultimately the $L_{mod}$ was chosen to be 10.2 cm.

The energy of the fast neutrons from the D-T reaction is above the threshold energy for several isotopes with non-zero (n, 2n) cross-sections. As such, a number of materials, such as beryllium, iron, tungsten, lead, etc. can be used to increase the epithermal neutrons flux via multiplication. Due to the ease of availability and manufacturing, and relatively low handling hazard, lead (Pb) was chosen as the choice of multiplier material. The multiplier material was placed directly in between the generator and the HDPE moderator to allow for the maximum neutron flux into the lead and for the (n, 2n) reactions to occur prior to neutrons entering the moderator.

A series of MCNP simulations with varying lead multiplier dimensions and placement was conducted to arrive at an optimized multiplying geometry. With Pb as a choice of multiplier, the multiplier, and moderator axial offsets were simulated. A 15 cm (i.e., $l_{mod}$) moderator positioned 5 cm in front of the target plane was held constant while a 10 cm wide multiplier of varying radial thickness was shifted axially relative to the target plane. After an optimization study, the optimal multiplier thickness was selected to be 7.5 cm. Due to limited material availability at time of moderator manufacturing, only a proof-of-concept multiplier geometry was built and experimentally tested. The as-built multiplier geometry differed from the optimized design in that the multiplier consisted of a 2.5 cm thick and 10 cm long annulus of lead around the generator target plane. The increased moderator offset was compensated for by adding an additional slab of 2.5 cm HDPE in front of the moderator assembly.

The epithermal flux from the HDPE-only moderator and HDPE + Pb moderator-multiplier geometries were compared by recording open beam ToF spectra using the GS20 detector and B4C + Cd shielding configurations (See Section 2.2 for details on detectors). Open beam refers to measurements with no target in the source-detector beam path. The average increase in the ~1-200 eV energy range was 48%. Table *1* below summarizes the relative increase in epithermal neutron flux for several energy ranges.

Table 1: Relative flux increase using HDPE + Pb design compared to HDPE-only moderator

| Approx. time range (µs) | Energy range (eV) | Relative increase (%) |
|---|---|---|
| 144.5 – 64.5 | 1 - 5 | 45 |
| 65.0 – 45.0 | 5 - 10 | 49 |
| 45.5 – 20.0 | 10 -50 | 51 |
| 20.5 – 14.5 | 50 - 100 | 49 |
| 15.0 – 10.0 | 100 - 200 | 46 |

*2.2    Epithermal Neutron Detector*

For the NRTA application discussed here, suitable detectors must be able to detect epithermal neutrons in the energy range of interest and have a timing resolution of better than 1-µs. Other practical considerations are commercial availability and ease of development. It will also be beneficial for the detector to have minimal sensitivity to gamma rays to limit possible gamma-related background contributions.

With these considerations in mind, the GS20 scintillator was chosen for neutron detection. GS20 is a glass scintillator doped with $^6$Li. In GS20 neutron detection is primarily performed via pulse-height analysis. With the lack of PSD capability, GS20 is limited in its ability to reject gammas. However, GS20 has been proven to function well in prior experimental studies [6] and thus it was chosen as the scintillator for the detection system. The detector was assembled by coupling a 50 mm diameter and 5 mm thick GS20 glass with a 3-inch diameter photo-multiplier tube.

*2.3    Shielding and Collimator*

NRTA background terms include off-axis background (neutrons that are detected after taking alternate routes to the on-axis neutron beam flight path) and on-axis background (room return neutrons that are scattered to the detector along the beam flight path). These background contributions were reduced with a neutron shield and collimator. Building upon a previous $B_4C$ shielding design [6], a modular shield was built from five $B_4C$ abrasive powder components housed in 3 mm thick aluminum walls. The assembled shield had a minimum $B_4C$ thickness of 7.62 cm in any direction, except the collimator opening. Step features were included at the interfaces of each of the five pieces to eliminate the potential for direct neutron streaming paths into the detector cavity. The room return neutrons are typically a result of fast neutrons scattering isotropically and thermalizing within the surrounding environment. These neutrons add to the background as they are detected through the front collimator opening. It is practically impossible to separate them in time without a generator capable of producing shorter pulses at much lower frequencies (e.g., 250 to 500 Hz). So, a 23 cm long collimator was integrated into the shielding to reduce on-axis scatter. The collimator had an 11.5 cm diameter opening to match the dimensions of the measurement targets.

The P-385 neutron generator can be operated at low frequencies, but with increasing pulse durations and overlapping neutron signals from prior pulses, which are referred to as wraparound neutrons.

Figure 5 shows the effect of the cadmium filter and B$_4$C shielding on neutron wraparound and off-axis background. These measurements were taken at 1000 Hz frequency with a duty factor of 5% (prior to generator pulse optimization) without a target present (i.e., open beam). At these settings, the neutron pulse is 50 μs wide. After this initial pulse, a factor of 2.5 reduction is observed in off-axis neutron counts. After approximately 500 μs, the two trends merge as the dominant neutron source is mostly in the form of room return from the front of the detector and through the collimator.

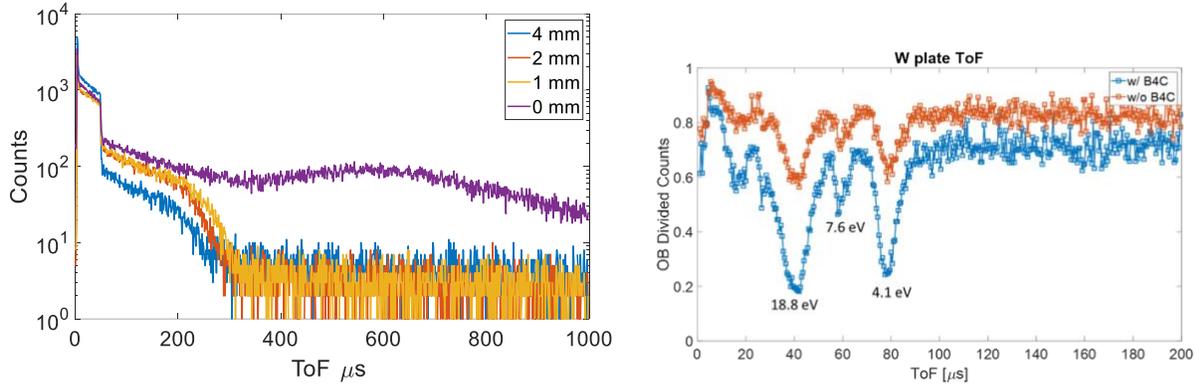

Figure 5: (left) Effects of cadmium thickness on off-axis background. The legend lists the cadmium thicknesses. (right) Comparison of shielded and unshielded ToF spectra, while using a 2mm on-axis Cd filter.

The cadmium-thickness optimization study showed that the cadmium plays a key role in suppressing wrap-around neutrons. Additionally, the use of B$_4$C showed a significant reduction in off-axis neutron contributions to the signal. Based on the above observations, it was decided that a 2 mm thick cadmium sheet is adequate for background reduction.

*2.4    Data Acquisition System*

The data acquisition system consisted of a CAEN desktop digitizer, DT5730SB, connected to a laptop. The digitizer consisted of eight channels operating at 500 mega-sample per second (MS/s) at 14-bit resolution. The analog pulses from the GS20 detector were readout using an Ortec 276 pre-amplifier module couple to the PMT. The analog pulses from the pre-amplifier were fed into the digitizer. A CAEN WDT5533E system was utilized to supply the PMT high voltage. To enable time tagging, the TTL logic pulse from the neutron generator was connected to the digitizer, which served as a time base for calculating the time of flight for each neutron event detected. The CoMPASS software provided by CAEN was utilized to set digitizer parameters and conduct data acquisition. Event-by-event (i.e., list mode) data was recorded to allow for neutron energy reconstruction (through the time-of-flight information) in post-processing.

The experimental setup developed under this effort was subsequently utilized for another study aimed at investigating the use of NRTA for nuclear safeguards application [11] .

## 3.0    MODELING, MEASUREMENTS & ANALYSIS

*3.1    Model*

Proof-of-concept measurements on special nuclear material (SNM) targets were conducted at the Low Scatter Facility (LSF) at PNNL. The larger and nearly clutter free environment allowed for measurements with very low neutron scatter. The room measures 16.5 meters in length, 10 meters in width, and 9 meters high. The room walls are 30-cm thick and made of concrete. The NRTA apparatus is on a raised platform, such that the neutron flight axis is 4 meters above the room floor and the detector is 5.7 meters from the nearest wall. A detailed MCNP [10] model of the room and the experimental setup, consisting of the source, moderator, multiplier, shield, and detector was created to gain a better understanding of the NRTA process and to simulate targets not available for experiments.

A D-T neutron source was included in the model. This source had an energy distribution with a mean of 14 MeV. The time dependence (distribution) of the neutron source pulse (Figure 3) was included in the model. Variance reduction (VR) was required to complete the modeling runs in a reasonable amount of time. The nested DXTRAN sphere VR technique was implemented in the MCNP model to increase the sampling efficiency for neutron detection [12]. The inner DXTRAN sphere circumscribed the GS20 detector, and the outer DXTRAN sphere enclosed the entire $B_4C$ shielding structure. This VR approach was highly effective, reducing the required computing time by a factor of approximately 300.

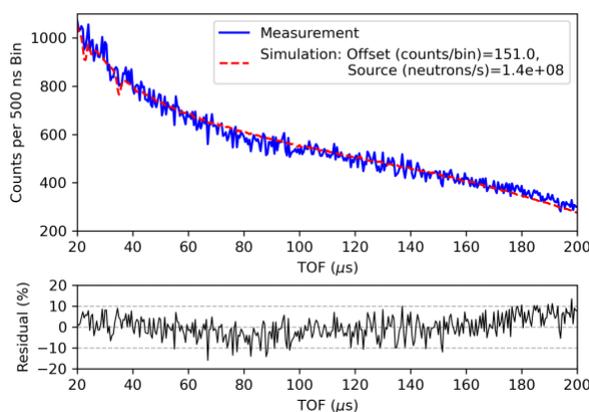

Figure 6. Comparison of the simulated and measurement ToF spectra for the open beam (i.e., no target sample) case.

Figure 6 shows the comparison of the modeled and measured open beam ToF spectra. A sufficient agreement between the modeled and measured spectra was achieved after applying the linear gain and offset correction to the modeled spectra. The benchmarked model is useful in simulating measurement scenarios otherwise not possible in a laboratory setting due to the unavailability of higher SNM mass, isotopic composition, and geometry configurations.

### 3.2 Proof-of-Concept Measurements

The proof-of-concept measurements were made using plate-type SNM materials to confirm the presence of the key resonance characteristic specific to the isotopes of interest. The targets for these measurements included: Depleted Uranium (DU), Highly Enriched Uranium (HEU), and Reactor-grade Plutonium (RGPu).

The experiments used various targets. The DU plate measuring 6 in. x. 6 in. x 0.13 in. was wrapped in a thin aluminized mylar foil and placed adjacent to the collimator opening. The HEU (93.5% 235U) target available for the experiment consisted of a circular foil weighing ~103 grams that was encased in polycarbonate. The polycarbonate thickness at the back and front of the HEU foil was estimated to be approximately 1 mm. The neutron scattering and absorption due to the hydrogen in the polycarbonate encasement results in blurring the time signature and overall reduction in detection efficiency. To obtain an accurate transmission for HEU, moderated open beam (or MOD OB) measurements were made, where a blank plastic holder identical to that of the polycarbonate around the HEU foil was placed in the beam path. The RGPu was enclosed inside a stainless enclosure along with other construction material that affected the overall transmission of neutrons and NRTA signature. The key interfering material inside the puck was tantalum. Open beam runs for Pu were made by placing in the beam only a steel and tantalum assembly (no Pu) of approximately the same thicknesses as in the original Pu casing to allow for an accurate calculation of the transmission spectrum. The ToF data with and without (i.e., open beam) the target was collected for 3600 seconds each.

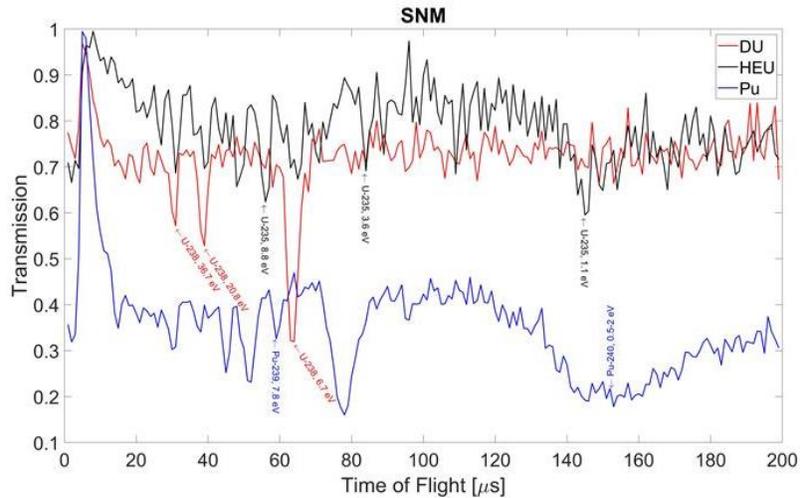

Figure 7: Measured ToF spectra of key SNM targets.

Figure 7 shows the measured ToF flight transmission spectra for the application-relevant target materials. The expected resonances in the usable range for the compact NRTA setup were distinctly visible. For the DU measurements, the resonances at 36.7 eV, 20.8 eV, and 6.7 eV characteristic to $^{238}$U were observed. Even with the presence of the low-Z attenuating material around the HEU, the characteristic $^{235}$U resonances at 8.8 eV, 3.6 eV, and the relatively broader resonance at 1.1 eV were visible. The key resonance at 7.8 eV for $^{239}$Pu was also observed. The broader resonances around 0.3 to 2 eV for $^{240}$Pu were also observed. Overall, the signatures specific to SNM were all observed in a 2-hour long measurement.

### 3.3    Analysis

The measured HEU and RGPu ToF spectra were fitted using the REFIT-2009 least square fitting code [13]. The analysis included the bin uncertainties. Figure 8 shows the measured and calculated ToF spectra for the HEU target. Initial isotopic abundances input to REFIT corresponded to an equal distribution of $^{234}$U/$^{235}$U/$^{236}$U/$^{238}$U and a non-resonant isotope in an attempt to remove bias towards a high or low enrichment. The residuals are given in terms of sigma. Table 2 gives the predicted abundances and total uranium thickness.

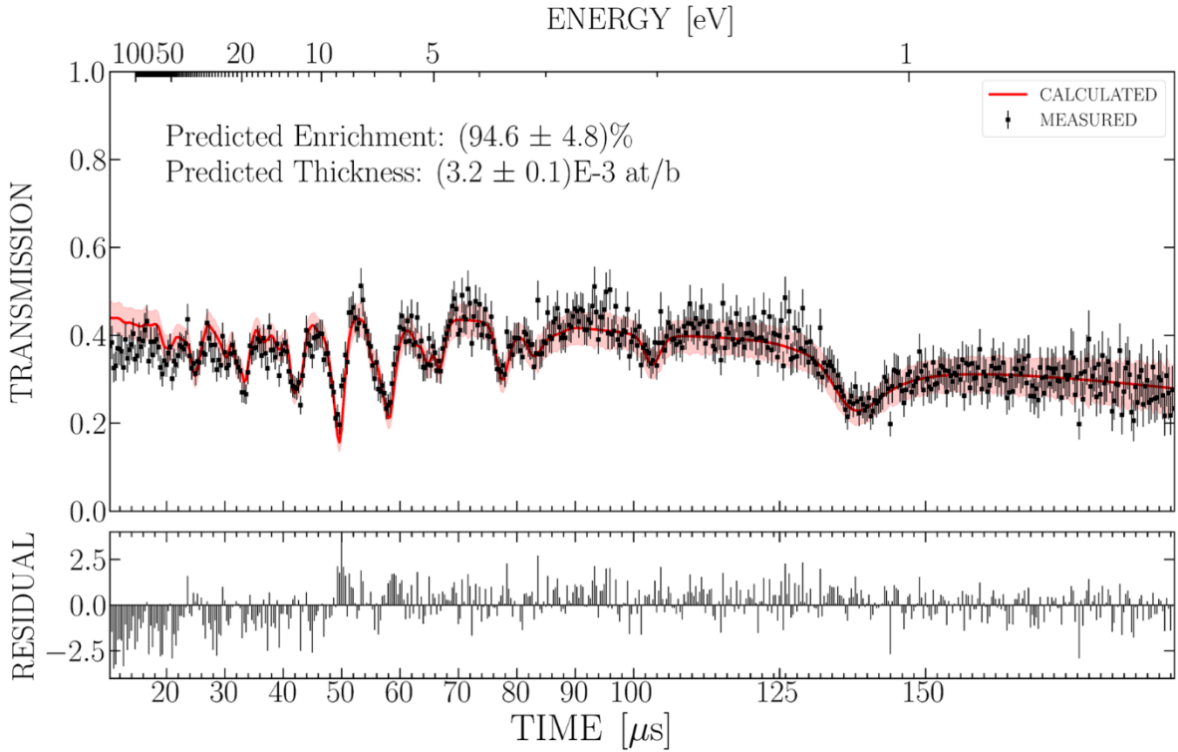

Figure 8: Measured neutron transmission spectrum with the REFIT fit (marked as calculated) for the HEU target

Table 2. True and REFIT-derived uranium abundances

| Isotope | REFIT-derived | True |
|---|---|---|
| $^{234}$U | (0.66 ± 0.15)% | 1.0% |
| $^{235}$U | (94.6 ± 4.8)% | 93.2% |
| $^{236}$U | (0.4 ± 0.2)% | 0.45% |
| $^{238}$U | (4.3 ± 0.8)% | 5.4% |

REFIT analysis estimated that the target had an enrichment of 94.6% ± 4.8% $^{235}$U, correctly identifying the material as HEU, though underpredicting the total uranium mass by approximately 25%. The HEU shielding was composed of several different plastic materials corresponding to an equivalent thickness of approximately 4 mm high-density polyethylene with an energy-independent neutron attenuation of 50%. The REFIT fit showed bias towards lower transmission values at longer times and higher transmission values at shorter times, potentially indicating the need for a slight correction to the time constant of the background term arising from the presence of hydrogenous material.

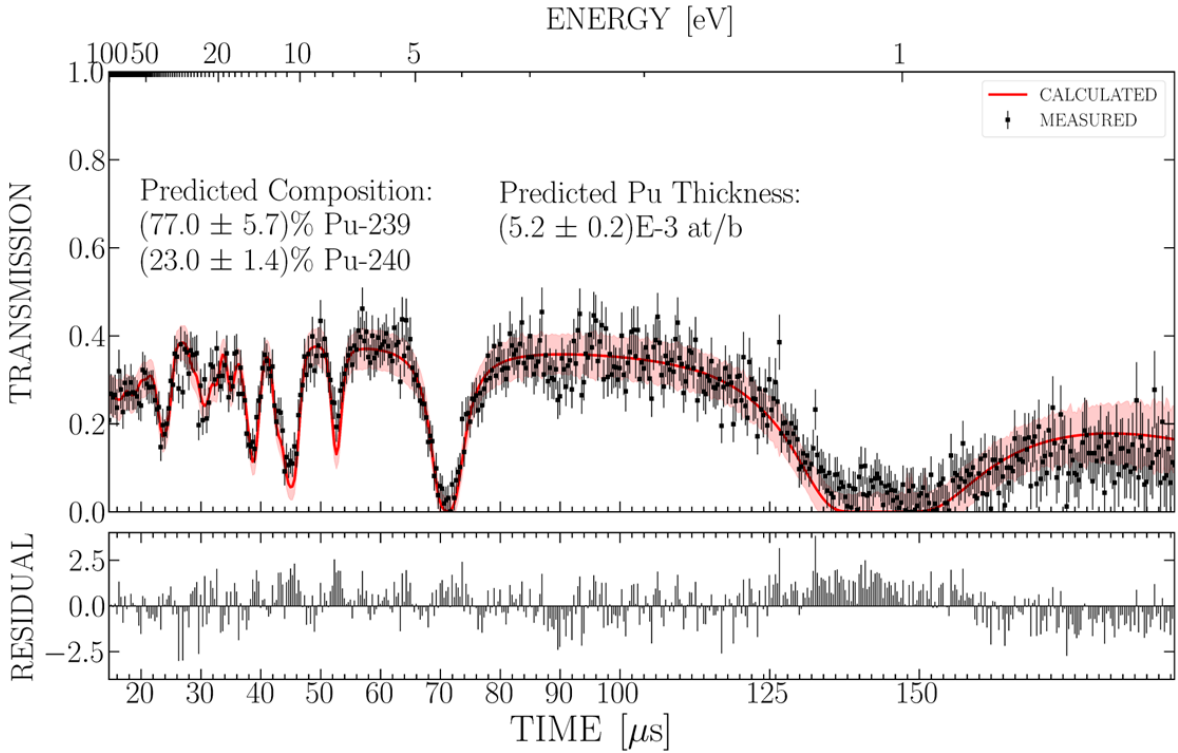

Figure 9: Measured neutron transmission spectrum for the RGPu target with the REFIT fit (in red).

Figure 9 shows the measured and fitted ToF spectra for the RGPu target. Table 3 shows the true and predicted abundance for the same target. REFIT estimated that the target had $^{239}$Pu and $^{240}$Pu content of (77.0 ± 5.7)% and (23.0 ± 1.4)%, correctly identifying the material as reactor-grade plutonium.

Table 3: True and REFIT-derived plutonium abundances

| Isotope | REFIT-derived | True |
|---|---|---|
| $^{239}$Pu | (77.0 ± 5.7)% | 80.7% |
| $^{240}$Pu | (23.0 ± 1.4)% | 18.5% |

## 4.0   CONCLUSION

A compact neutron time of flight measurement system has been built and tested to enable NRTA measurements with a short flight distance of 2 meters. The system elements, consisting of a pulsed DT neutron source, detector, and data acquisition hardware, have been optimized to achieve the energy resolution needed to observe isotopic-specific resonance attenuation dips in the epithermal range. The proof-of-concept measurements of SNM targets further showed that the physics-based observables, i.e., the characteristic resonances, were confirmed using two-hour long measurements. An accurate model of the full-scale system was developed and verified against measurements. The model was reliable in conducting investigational studies otherwise not possible in a laboratory setting, for example, varying SNM mass and geometry configurations.

Potential analysis techniques were investigated to assess the applicability of NRTA for arms control verification purposes. These included a simple region-of-interest based metric and a more advanced multi-parameter least-squares ToF spectral fitting technique (REFIT). While the simple metric proved to be less reliable due to the high uncertainty resulting from low counts and non-ideal source geometry, it may still be applicable in certain cases where the geometry is more favorable and relatively free of scatter components. The REFIT analysis showed that the least-squares fitting approach is reliable in extracting target attributes for both the measured and simulated ToF spectra. Overall, during this early-stage investigation, the measurements, modeling, and analysis show that NRTA holds the potential to be a viable verification and confirmation tool for arms control applications. Additional experimental evaluation on application-specific geometries and modeling is needed to further refine the method and determine system limitations (material thickness, for example).

Further refinements to the system hardware, analysis methods, and targeted measurements can improve the overall methods and bring NRTA a step closer to the envisioned application space. On the hardware front, a neutron generator with a higher output and narrower pulse structure can improve energy resolution. A more compact detection system can be constructed via the use of silicon photomultiplier-based scintillator readout. With a more compact detector system, the surrounding shielding can be reduced to nearly half the size and weight. Higher fidelity measurements on representative objects, geometries, and varying SNM mass can be made to strengthen the technique for potential treaty verification applications.

## ACKNOWLEDGEMENTS

This work is supported by the US Dept. of Energy, National Nuclear Security Administration, Office of Defense Nuclear Nonproliferation Research and Development. The work by A. Danagoulian, F. Naqvi, and E. Klein was supported by DOE NNSA NA-221 award DE-NA0003920.

# References

1. Seager, K., et al. *Trusted radiation identification system*. in *Proceedings of the 42nd Annual INMM Meeting*. 2001.
2. Mitchell, D.J. and K.M. Tolk, *Trusted radiation attribute demonstration system*. 2000, Sandia National Labs.
3. Merkle, P.B., et al. *Next Generation Trusted Radiation Identification System*. in *Proceedings of 51st Annual INMM Meeting. Baltimore, MD*. 2010.
4. Reilly, D., et al., *Passive nondestructive assay of nuclear materials*. 1991, Nuclear Regulatory Commission, Washington, DC (United States). Office of ….
5. Engel, E.M., E.A. Klein, and A. Danagoulian, *Feasibility study of a compact neutron resonance transmission analysis instrument.* AIP Advances, 2020. **10**(1): p. 015051.
6. Klein, E.A., et al., *Neutron-Resonance Transmission Analysis with a Compact Deuterium-Tritium Neutron Generator.* Physical Review Applied, 2021. **15**(5): p. 054026.
7. Engel, E.M. and A. Danagoulian, *A physically cryptographic warhead verification system using neutron induced nuclear resonances.* Nature communications, 2019. **10**(1): p. 1-10.
8. Brown, D.A., et al., *ENDF/B-VIII. 0: the 8th major release of the nuclear reaction data library with CIELO-project cross sections, new standards and thermal scattering data.* Nuclear Data Sheets, 2018. **148**: p. 1-142.
9. Moss, C.E., et al., *Survey of Neutron Generators for Active Interrogation*. 2017, Los Alamos National Lab.(LANL), Los Alamos, NM (United States).
10. Werner, C.J., et al., *MCNP version 6.2 release notes*. 2018, Los Alamos National Lab.(LANL), Los Alamos, NM (United States).
11. McDonald, B.S., et al., *Neutron resonance transmission analysis prototype system for thorium fuel cycle safeguards.* Nuclear Instruments and Methods in Physics Research Section A: Accelerators, Spectrometers, Detectors and Associated Equipment, 2024. **1062**: p. 169148.
12. Booth, T., K. Kelley, and S. McCready, *Monte Carlo variance reduction using nested dxtran spheres.* Nuclear technology, 2009. **168**(3): p. 765-767.
13. Moxon, M., T. Ware, and C. Dean, *REFIT-2009 A Least-Square Fitting Program for Resonance Analysis of Neutron Transmission.* Capture, Fission and Scattering Data Users' Guide for REFIT-2009-10 (UKNSFP243, 2010), 2010.

# APPENDIX A – Moderator Design

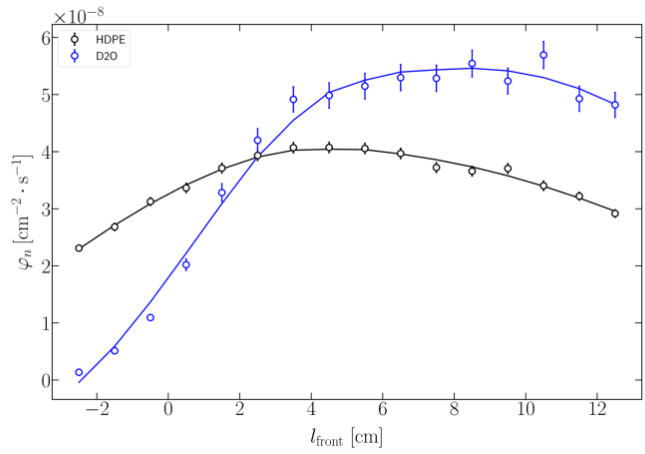

Fig. A. 1: Epithermal neutron flux from HPDE and D₂O as a function of L$_{front}$.

Fig. A. 1 shows the simulated epithermal neutron flux output from HPDE and D₂O moderators as a function of moderator placement (i.e., $l_{front}$). The output from D₂O is larger than that of HDPE. However, D₂O requires an additional 5 cm of moderator material. The larger moderator length means increased moderation delay resulting from neutrons interacting in the material for a longer time and uncertainty in measured neutron energy. For all neutron energies, moderator delay distance for HDPE and D₂O was calculated to be (1.910 ± 0.003) cm and (10.48 ± 0.02) cm, respectively.

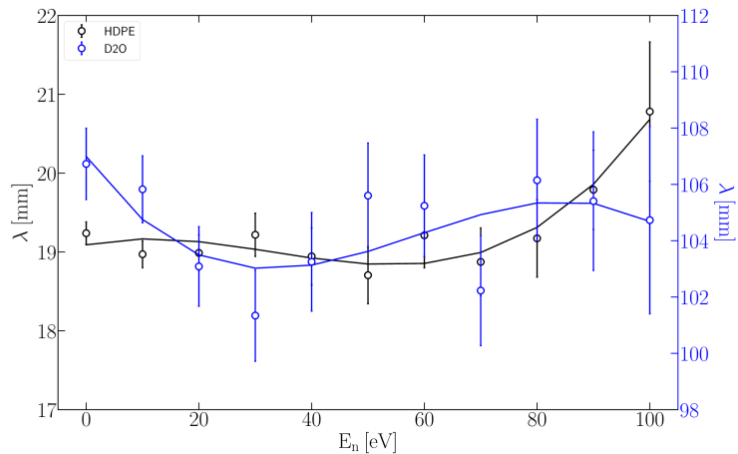

Fig. A. 2: Moderation delay distance in HDPE and D2O as a function of neutron energy

The moderation delay distance as a function of neutron energy is shown in Fig. A. 2. Note the approximately 6 times higher magnitude on the y-axis for D₂O than that of HDPE. Except for a minor increase beyond 80 keV, the delay distance is largely independent of energy. Therefore, the moderation delay distance can be factored as a constant distance to the true ToF flight distance: $d_{total} = d_{mod} +$

$\sqrt{d_{rad}^2 + d_{ToF}^2}$. The uncertainty in delay distance, which contributes to the overall system resolution, across all energies is $d_{total}/\sqrt{3}$. For HDPE and D₂O, the mean moderation distance is (1.10 ± 0.01) cm and (6.05 ± 0.05) cm. For a 2-meter ToF distance, this corresponds to a relative energy resolution of 1% and 6% for HDPE and D₂O, respectively. Given the significantly higher energy resolution coupled with only a minor decrease in neutron flux, it is evident that HDPE moderator is more optimal for portable NRTA setup using short ToF distances.

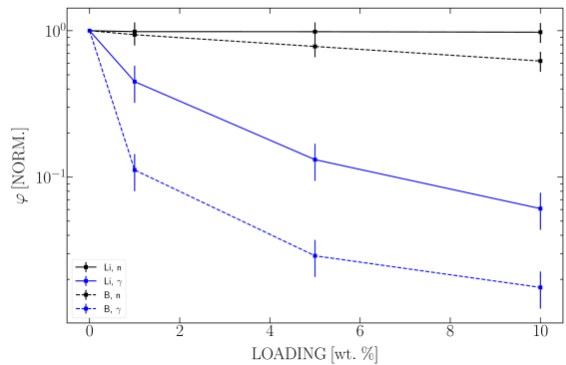

Fig. A. 3: Relative gamma and neutron flux as a function of neutron poison loading concertation in HDPE

Fig. A. 3 shows gamma and neutron flux relative to HDPE as a function of loading concentrations.

Fig. A. 4 shows the effect of the moderator dimensions on epithermal neutron flux. The flux plateau starts at 5 cm of moderator thickness in the front with respect to the target plane. Additional moderator material beyond this thickness leads to increased thermalizations and therefore produces a lower epithermal flux. Axial epithermal flux increases as additional moderator material is added at the back of the target plane. This is likely due to increased reflection of fast neutrons towards the front end of the moderator. The plateauing of the flux starts at 6 cm of moderator at the back. Per the trends shown in Fig. A. 4, the optimal moderator thickness is 5 cm in the front and 6 cm at the back relative to the target plane, giving the overall thickness of 11 cm. Readily available, on-hand HDPE consisted of 2.54-cm thick slabs. To reduce the weight and for ease fabrication, it was decided that four 2.54 cm sheets are adequate as the reduction of the moderator material by 1 cm from the back results in an insignificant decrease in neutron flux. Therefore, the optimal HDPE moderator thickness was determined to be ~10.2 cm.

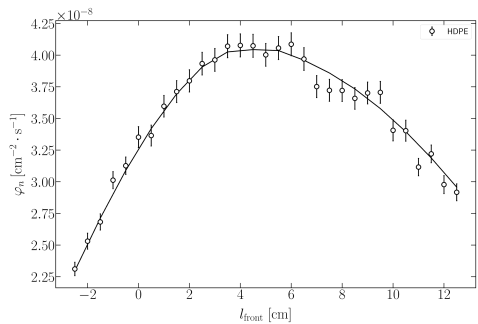

(a) front

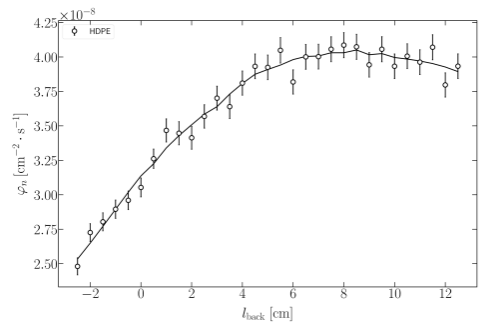

(b) back.

Fig. A. 4: Epithermal neutron flux as a function of moderator front and back dimensions.

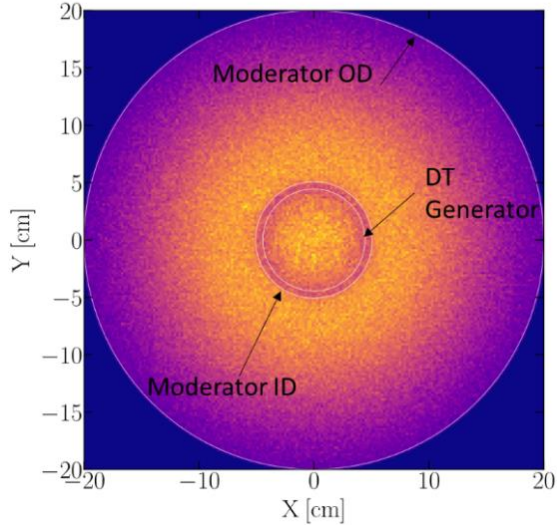

Fig. A. 5: 2D distribution of epithermal flux exiting the front of the moderator and DT- tube.

The emission of the epithermal neutrons from the front of the moderator is concentrated at the center of the source assembly. The flux reduces towards the circumference of the D-T tube. Most of the neutrons start to moderate and exit the front face near the inner radius of the moderator. The flux decreases towards the edge of the moderator. This trend can be qualitatively observed in Fig. A. 5, which shows the mesh tally of epithermal neutron flux exiting the front of the source assembly.

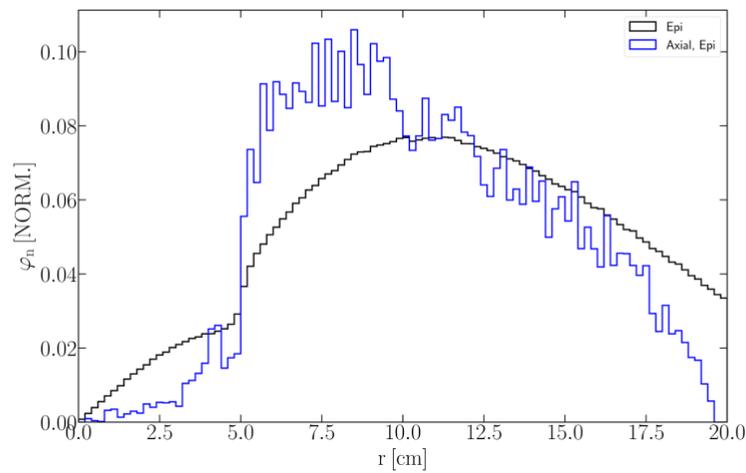

Fig. A. 6: Probability distribution function of total and axial epithermal flux as a function of radial distance from ToF axis

Considering only the axial epithermal flux, only 4% of it is emitted from the 4" diameter of the DT tube. Fig. A. 6 shows the probability and cumulative distribution function for a 15 cm (20 cm outer diameter with a 10 cm circular cutout in the center for the generator tube) radially wide HDPE moderator. A significantly smaller fraction of the axial epithermal neutron flux is emitted from within the diameter of the neutron generator tube than for all epithermal neutrons. The median radial distance is 8.5 cm and 85% of axial epithermal neutrons were emitted within a radial distance of 15.0 cm from the ToF axis. The axial epithermal flux decreases to essentially zero at 20 cm radial distance. Due to the limited availability of the material in desired dimensions at the time of moderator fabrication, the final moderator dimensions for the HDPE-only design were 28 cm outer diameter, 10.5 cm inner diameter, and 10 cm thick (axial along the generator axis). Note that the collimator opening limits the number of neutrons originating near the edge of the moderator from making it to the detector. Therefore, the reduction in moderator radius compared to the optimized radius does not result in a significant reduction in usable epithermal neutron flux. The as-built dimensions vary slightly compared to the optimized dimensions stated earlier due to the dimensions of the raw material available on hand.

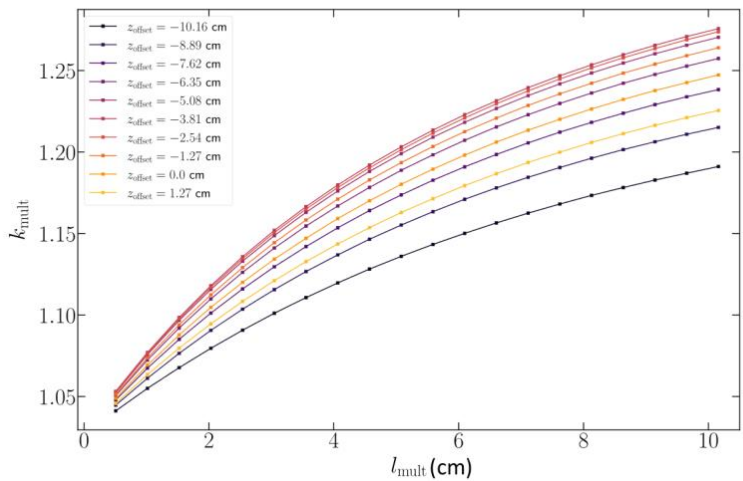

Fig. A. 7: Neutron multiplication as a function of Pb multiplier radial widths

Fig. A. 7 depicts the effect of offset and multiplier width. The multiplication increases as the multiplier is offset backwards.

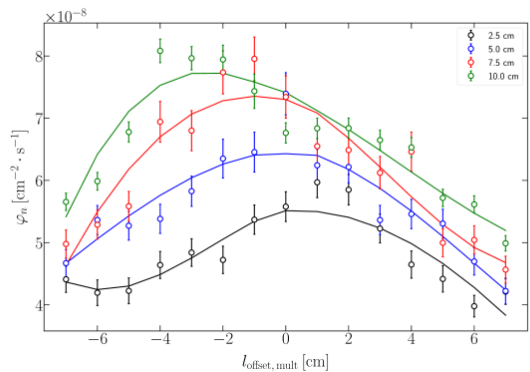

(a) Multiplier Offset

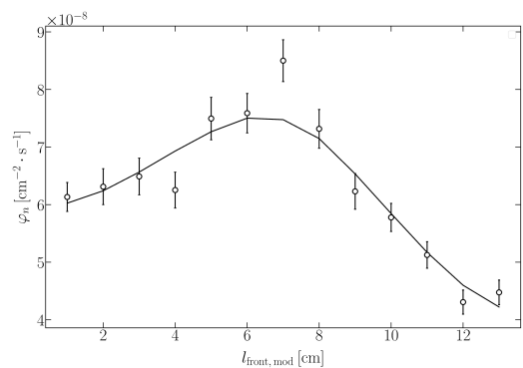

(b) Moderator Offset

Fig. A. 8: Effect of moderator and multiplier offsets on axial epithermal flux for various multiplier radial thicknesses.

Fig. A. 8a shows the effect of multiplier offset with respect to the target plane on epithermal flux for varying multiplier radial thicknesses ranging from 2.5 – 10 cm. As the multiplier width increases, the optimal multiplier offset shifts further backward relative to the front of the moderator. Given the trends in Fig. A. 8a, the optimal multiplier thickness was selected to be 7.5 cm. For a fixed multiplier position at its optimal value, the axial epithermal flux output is sensitive to the moderator offset. Compared to the optimal position of 4 – 6 cm the moderator without the multiplier (Fig. A. 4), Fig. A. 8b shows that the optimal position of the moderator is 7 cm from the front target plane with the multiplier present. Once again, due to schedule constraints and limited material availability at time of moderator manufacturing, only a proof-of-concept multiplier geometry was built and experimentally tested.

Fig. A. 9 shows the mean moderator delay distance for the final source geometry. The moderator delay distance is smaller than which was seen for the moderator-only and moderator with multiplier designs likely due to the removal of additional HDPE material from behind the multiplier. Therefore, there is no significant trade-off between the increased neutron flux using the multiplier and the system resolution.

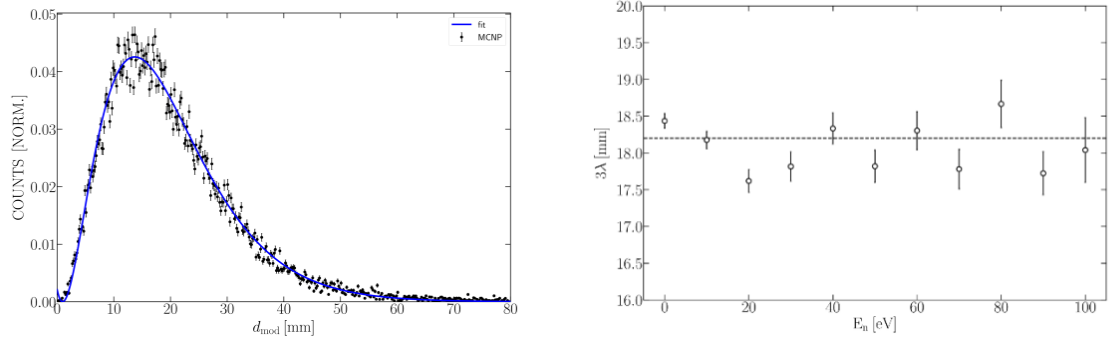

(a) Mean delay distance distribution for epithermal neutrons with chi-square fit.

(b) Mean delay time distance as a function of incident neutron energy.

Fig. A. 9: Moderator delay distribution for the final source assembly

Fig. A. 10 shows epithermal neutron counts as a function of time of flight. The curve in blue represents counts vs. time trend obtained using an HDPE-only moderator. The curve in orange represents the same using the multiplying moderator (i.e., HDPE + Pb). The y-axis on the right represents the ratio of the two trends. The average increase of epithermal neutrons was 48%.

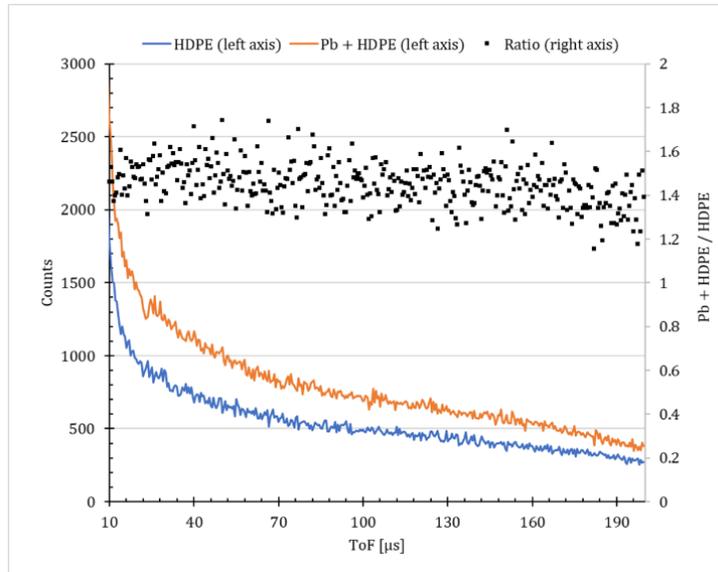

Fig. A. 10: Comparison of HDPE-only and HDPE + Pb geometries.